# Copper sulfide nanosheets with shape-tunable plasmonic properties in the NIR

Rostyslav Lesyuk[1,2], Eugen Klein[1], Iryna Yaremchuk[3], Christian Klinke[1,4,*]

[1] *Institute of Physical Chemistry, University of Hamburg, Martin-Luther-King-Platz 6, 20146 Hamburg, Germany*
[2] *Pidstryhach Institute for applied problems of mechanics and mathematics of NAS of Ukraine, Naukowa str. 3b, 79060 Lviv, Ukraine*
[3] *Department of Photonics, Lviv Polytechnic National University, S. Bandera Str. 12, Lviv 79013, Ukraine*
[4] *Department of Chemistry, Swansea University - Singleton Park, Swansea SA2 8PP, United Kingdom*

**\*** Corresponding author: christian.klinke@swansea.ac.uk

**Graphical abstract**

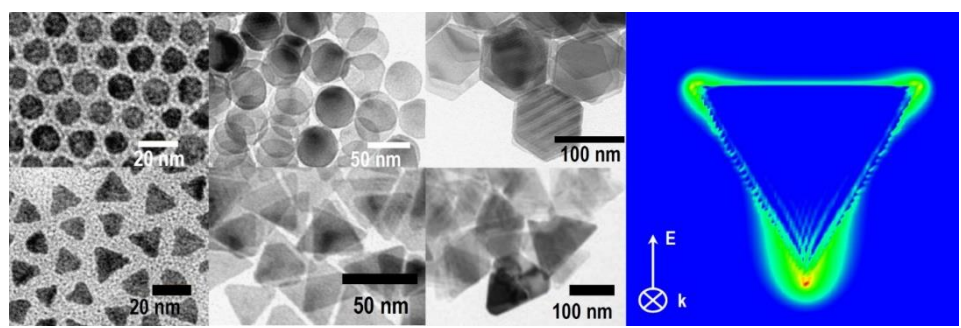

**Abstract**

2D copper sulfide nanocrystals are promising building blocks for plasmonic materials in the near-infrared (NIR) spectral region. We demonstrate precise shape and size control (hexagonal/triangle) of colloidal plasmonic copper sulfide (covellite) nano-prisms simply by tuning the precursors concentration without introduction of additional ligands. The ultra-thin 2D nanocrystals possess sizes between 13 and 100 nm and triangular or hexagonal shapes. We also demonstrate CuS nanosheets (NSs) with lateral sizes up to 2 microns using a syringe pump. Based on the experimental findings and DFT simulations we propose a qualitative and quantitative mechanism for the formation of different shapes. The analysis of the spectral features in the NIR of synthesized CuS nanocrystals has been performed in respect to the shape and the size of particles by the discrete dipole approximation method and the Drude-Sommerfeld theory.

**Introduction**

Copper sulfide ($Cu_xS$) was an extensively studied material in respect to solar cells in the 60s and 70s of the last century [1]. It is an abundant compound with almost a dozen of phases which crystallize with stoichiometries ranging from x=1 (covellite) to x=2 (chalcocite) and with band gaps covering the 1.3–2.4 eV range [2-4]. Its application in the field of photovoltaics was hindered through the relatively high degradation rate of the $Cu_xS$/CdS interface (CdS served as a window layer for the p-n junction) and the $Cu_xS$ absorber [5]. However, during the last decade the attractiveness of the family of copper chalcogenides as cheap low-toxic materials gained attention after several protocols were published for nanostuctured copper chalcogenide synthesis by wet chemical methods [6-13] as promising fabrication method for solution-processable devices like flexible conductive films [11], sensors, solar cells [10, 14, 15], electrodes for the Li-ion batteries [16] or even moieties for cancer photothermal therapy (CuSe [17]).

The plasmonic activity of copper sulfides deserves special attention [9, 18-22]. It usually arises from the non-stoichiometric composition with a deficiency in copper, which leads to a partial filling of sulfur 3p orbitals, which in turn leads to the formation of vacancies [23]. Latter contribute to free holes, producing high electrical conductivity in pure, degenerately-doped (self-doped) copper sulfides. With the decreasing amount of copper in the compound, the amount of charge-carriers increases, reaching substantial concentrations of about $10^{20}$ to $10^{21}$ cm$^{-3}$ [20]. The crystal phase called covellite (CuS) supplements the copper sulfide family with unique physical properties like high holes concentration (up to $10^{22}$ cm$^{-3}$ [19]) based on the specific crystal structure. Additionally the high degree of hole delocalization is inherent to the covellite phase [24], as well as distinct anisotropic metallic conductivity [25]. Thus, this semiconductor shows a highly pronounced localized surface plasmon resonant (LSPR) absorption peak appearing in the NIR [26] (in the second biological spectral window, 1000–1350 nm [27, 28]) in comparison to gold, silver and copper, which are located in the visible range. Control over the position, oscillator strength and the width of the LSPR peak can be ensured by the size, shape [19] and surface chemistry adjustment of the nanostructure [29, 30]. Under specific conditions, CuS nanoparticles crystallize as 2D nanocrystals (NCs) namely as nanodisks (or nanoprisms) [12] according to the lamellar growth mechanism [8] and due to the anisotropic crystal structure (hexagonal, P63/mmc space group). According to recent theoretical studies, copper ions bind tetrahedrally and trigonally to sulfur ions such as in summary $[(Cu)_2]^{3+}$ and $Cu^+$ to $S_2^{2-}$ and $S^{2-}$, whereas S–S bonds are covalent [24, 31]. For the 2D growth, the lateral growth directions are perpendicular to the z-axis of the hexagonal crystal, and the preferable cleavage plane would be the one containing the S–Cu bonds [31, 32].

Shape and size control over the CuS NCs could provide a useful instrument for tuning of fundamental optical and plasmonic properties in the NIR region based on the geometry of the particles. This includes the tuning of the LSPR peak position and the electrical field distribution over the particle and in its vicinity. It has been repeatedly shown that the near-field distribution becomes strongly non-homogeneous with hot-spots around the corners reported for silver, gold and copper triangular nanoprisms [33, 34]. This in turn would allow the fabrication of very sensitive, spatially enhanced antennas and sensors [35] employing the triangular shape, or it would open the possibility to introduce CuS NCs as energy-mediated species into spectrally enhanced solar cells and spectral converters. CuS 2D NCs are known to crystallize usually with the hexangular, triangular or truncated triangular shape. Usually, the morphological yield of the wet chemical CuS synthesis contains all shapes at once, e.g. Ref. [36]. Gradual shape transformations to triangles has been reported in the presence of halogen ions in a thorough studies of Tao's group for the wet synthesis of CuS [37] or depending on the organic ligand concentration, namely trioctylphosphine oxide (TOPO) for $Cu_{2-x}S$ NCs shown by group of Donega [38]. In this letter, we demonstrate a simple and effective approach for the shape control of CuS nanoprisms in the size range between 13 and 100+ nm based on the precursors reactivity balance tuning. We propose a mechanism for the shape transformations between triangular and hexangular 2D nanoprisms supported by DFT simulations of the ligand adsorption energy on relevant NC facets. We also discuss the optical properties of our system in terms of shape influence on the plasmon resonances and absorption spectra.

**Experimentals**

*Chemicals.* Copper (I) iodide (CuI, 99.5%), Copper (I) acetate (CuOAc, 97%), oleylamine (OLAM, 70%), sulfur powder (99.98 %), tetrachloroethylene (TCE, spectroscopy grade Uvasol) were purchased from Sigma-Aldrich; toluene (99.5 %), methanol (>99.9%) were purchased from VWR, acetone (99%) was purchased from Th. Geyer. Chemicals were used as received without additional purification.

*Synthesis of CuS NCs.* The CuS NCs were synthesized according to a procedure published in ref. [12] with modifications using a standard Schlenk-line technique. Briefly, 0.1 to 0.8 mmol of CuI or CuOAc were mixed with oleylamine (OAm, 20–30 mL, 70% grade, Sigma) and added to a 25 mL three-neck flask, vigorously stirred and degassed by applying a vacuum for 30 min. Then, the flask was filled with nitrogen and gradually heated up. Then 2 mL of S/OAm solution (0.5 M) previously degassed and purged with nitrogen, was hot-injected at 120–150$^o$C.

During the reaction, aliquots were taken for spectroscopic characterization and TEM analysis. Gradual addition of precursors (0.02–0.08 M solution of CuI or CuOAc in OAm and 0.2 M solution of sulfur in OAm) was carried out with the velocity of 0.04–0.085 mL/min with separate syringes. After cooling down to room temperature, the reaction mixture was diluted with toluene followed by the purification of the nanocrystals. They were precipitated by adding a methanol/acetone mixture (1:1 vv) followed by centrifugation. The precipitate was resuspended in toluene and the washing procedure was repeated twice. For the spectroscopic characterization, the samples were redispersed in TCE.

*DFT simulations.* The software package CP2K [39] with the PADE LDA functional, the DZVP basis set, and a corresponding GTH-PADE potential were used for the evaluation of adsorption energies of ligands on specific facets. The crystal geometry had been kept fixed to the experimental values for the CuS covellite crystal and the ligands were free to relax by geometry optimization. An individual CuS nanocrystal with 230 Cu and 230 S atoms and the respective ligand molecules were simulated with periodic boundary conditions where the box dimensions are sufficiently large to avoid interaction between virtual neighboring molecular structures.

**Characterization**

Transmission electron microscope (TEM) images and selected area electron diffraction (SAED) patterns were obtained on a JEOL-1011 (100 kV). Samples for the TEM analysis were prepared by drop-casting of 10 µL diluted in toluene dispersion of nanocrystals onto carbon-coated copper or titanium TEM grids followed by solvent evaporation. The high resolution (HR) TEM images were obtained with a Philips CM 300 UT microscope operated at 200 kV. X-ray diffraction (XRD) measurements were performed with a Philips X'Pert System with Bragg-Brentano geometry, equipped with a copper anode (Kα X-ray wavelength of 0.154 nm). Samples were prepared by drop-casting of the colloidal nanocrystal solution onto silicon wafer substrates (<911> or <711> cut) with subsequent solvent evaporation. Atomic force microscope (AFM) measurements were performed on a JPK Instruments system (JPK Nano Wizard 3) in intermittent contact mode. UV-VIS-NIR absorption spectra were obtained with a Cary 5000 spectrophotometer equipped with an integration sphere.

**Results and Discussion**

*Size and shape transformations and control.* We employed the hot-injection approach of Wu *et al.* [12] where the S-OLAM mixture is introduced to the copper (I) precursor (CuI) in OAm at moderate temperatures without any further co-ligands. For the thickness adjustment we employed a DPE/OLAM mixture as solvent. To understand the role of the iodide in the synthesis coming from the precursor we alternatively to the CuI-based synthesis performed halogen-free reactions based on copper acetate and copper oleate as precursors.

The balance of the reactivity of participating components plays an important role [40, 41]. In our case, the copper cation being a soft Lewis acid should be appropriately coordinated by certain soft Lewis bases (iodide anions) to provide sufficiently low reactivity compared to elemental sulfur in OAm in order to form stoichiometric $Cu_1S$. Although the stoichiometric window for the formation of the pure covellite phase is either narrow, we were able to study the influence of the Cu:S ratio on the morphology of the nanostructures taking several intermediate aliquots during the synthesis procedure.

Starting with a 0.8:1 (Cu:S) ratio we observe nucleation of irregular round covellite nanodisks which slowly ripen retaining the shape (Figure 1 A, B; Corresponding XRD pattern is shown in Fig. 1 C). Few triangular platelets are also observed. When the Cu:S ratio is decreased to 0.4:1, the triangular shape fraction increases (Fig. 1 D). Under these conditions, the system typically undergoes three phases shown in Fig. 1 D–I: 1) nucleation (phase 1, kinetic growth mode); 2) size-shape-transformation phase, characterized by thermodynamic growth and ripening; 3) anisotropic growth at elevated temperatures.

During this process, in the first phase, initial nanoparticles crystallize with irregular shape with a tendency to triangles followed by anisotropic growth and shape focusing (phase 1, start), observed also by Hsu *et al.* [37] for a CuS one-pot synthesis. In parallel, we observe a growth of hexagonal platelets despite the presence of halogen ions (iodide) in the flask. With proceeding reaction time (5–120 minutes), the monomer supply rate for the growth decreases and further ripening occurs in expense of smaller triangles, which start to dissolve (Fig. 1 D–G). Hexagonal particles ripen further while full dissolution of smaller ones takes place. Eventually, the growth of hexagonal nanoprisms reaches saturation and stops. Monitoring of the synthesis revealed the ability of CuS NCs to grow anisotropically to nanosheets with 1–2 micron lateral size at elevated temperatures in the range of 140–180°C, if an additional amount of precursors is supplied to the flask like it was shown for CdSe nanoplatelets [42]. The micron-sized CuS NSs possess thicknesses of about 10 to 40 nm depending on the synthesis conditions (Supporting information (SI), Fig. S1 for AFM image). The XRD pattern of as-prepared $Cu_xS$ NCs can be found in Fig. 1, C. XRD reflexes of different copper sulfide crystal phases appear at similar values, and some of them coincide like in the case of covellite, yarrowite, talnakhite and digenite for example. However, the covellite phase can be distinguished by simultaneous appearance of peculiar reflexes at 10° (002), 27.2° (100), 27.7° (101), 39.5° (102) and 48° (110). In general, the observed XRD patterns match the hexagonal structure of covellite well and excludes the presence of other phases (see Fig. S2 for the XRD of different shapes of CuS). The intensity of the reflexes for 2D NCs is influenced by preferential orientation of nanoprisms on the substrate and will be discussed later in this section.

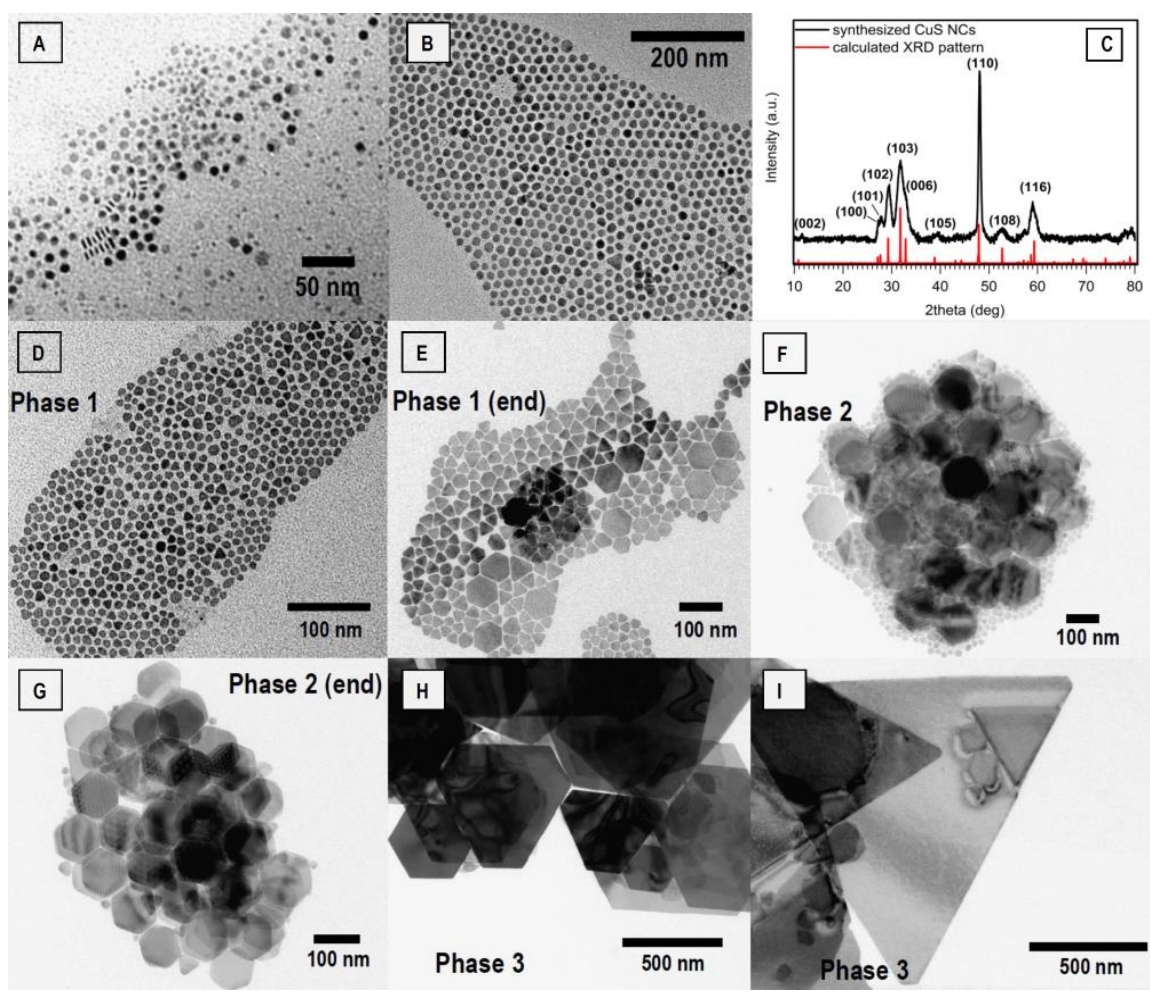

**Figure 1.** Temporal evolution of the $Cu_xS$ 2D system under different Cu:S precursor ratios: (A,B) 0.8:1 (1 min, 60 min); (C) XRD pattern of $Cu_xS$ NCs. Reference: calculated profile based on the crystal structure for covellite from Ref. [43]; (D,E,F,G) 0.4:1 (1 min, 30 min, 60 min, 120 min). Reaction temperature 120 °C; (H,I) growth at elevated temperature with precursor addition (160° C for 60 min and 150 °C for 16 h DPE:OLAM 1:1 vv).

We noticed, that a further decrease of the Cu:S ratio promotes the formation and growth of faceted triangles with sizes in the range between 30 and 50 nm (Fig. 2) and thickness of approximately 3 to 5 nm (Fig. S2 B). This indicates that the formation of triangles is conditioned by two factors – a kinetic mode of the growth and balancing of the reactivity of Cu and S (synthesis under sulfur excess).

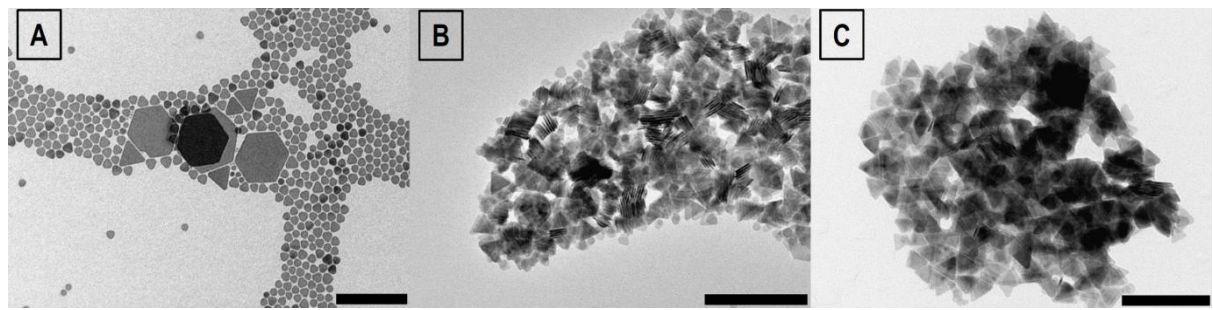

**Figure 2.** TEM images of CuS NCs after 5 min of synthesis by 120°C with Cu:S ratios 0.4:1 (A), 0.2:1 (B) and 0.1:1 (C). Scale bar 200 nm.

One of additional factors for the shape control is the amount of non-coordinating solvent (e.g. DPE), introduced to the synthesis in parallel to the OAm. An increase of the overall amount of solvent promotes nucleation of a fewer amount of bigger particles and allows retaining the system in the kinetic growth mode for a longer period due to higher available monomer buffer. Dilution of OAm with up to 50% of DPE provides the advantage for monomers coming to the surface of NCs in competition with OAm. Thus, triangular NCs can be observed as dominant shape with lateral dimensions even up to 200 nm upon addition of monomers using a syringe pump (SI, Fig. S3).

The influence of halogens on the shape of NCs has been frequently reported both for semiconductors [44-48], in particular for copper sulfide [37, 38], and for metals [49, 50]. In many cases, halogens selectively adsorb on specific facets promoting shape transformations [45, 51]. In our case, although the copper/iodide ratio was identical in all above-mentioned syntheses, the halogen ions could have some impact on the shape of NCs. We studied the structure of synthesized NCs by selected-area electron diffraction (SAED). We synthesized triangular and hexagonal nanosheets in the submicron range and studied the SAED patterns of individual particles to assign the crystalline orientation in respect to their shape. Exposing the sheets from the [001] zone axis with the electron beam, we were able to assign the side facets of the triangular NCs to (100), (010) and ($1\bar{1}0$), which have identical atomic cross-section and thus equal surface energies (Fig. 3) similar to observations of Du et al. [36]. This feature follows from the symmetry of the crystal which belongs to the P63/mmc space group. The appearance of the hexagonal shape is conjugated with the development of a facet between (010) and (100) which is the ($1\bar{1}0$) (Figure 3). It follows that all side facets of a CuS hexagonal NS are identical in their structure and stoichiometry and that they are either copper-rich or sulfur terminated. This excludes the scenario of preferred lateral growth by the selective passivation, since all six directions are energetically equivalent. Since at the end of the syntheses we get regular triangles, truncated triangles or regular hexagons, in general regular or symmetric shapes as our major morphological product, we can assume that in phase 1 exactly triangles are the initial faceted shape as a basis, and during the ripening process they develop to truncated triangles and hexagons.

The appearance of triangles during the hot-injection synthesis is not trivial to interpret, and could be rationalized by the fact that during the first moments of the precursor mixing the local concentration of copper cations and especially sulfur (which is injected) is at its maximum assuring prompt nucleation and growth in the kinetic regime. During the growth of NCs, OAm as the solvent and ligand passivates the copper surface cations due to the nucleophilic character of the amine group. We simulate the interaction of the amine and carboxylic groups with different facets using a DFT-based approach. In particular, we see that the adsorption energy for the Cu-rich (100) facet is essentially larger than that for (110) (Table 1). As a consequence, the (100) facets grow slower. This difference constitutes the growth of facets in triangles and hexagons with (100)-like edge facets. Atomistic 3D models for CuS NC are represented in Fig. 4 A, B.

**Table 1.** Adsorption energies of amine and carboxyl groups on different side facets of CuS nanocrystal, simulated using DFT.

|  | Adsorption energy, eV | | | |
|---|---|---|---|---|
|  | {100} #1 (Cu-rich) | {100} #2 (Cu-rich) | {100} #3 (S-terminated) | {110} |
| C2 Alkylamine ($CH_3–CH_2–NH_2$) | 1.66 | 2.61 | 0.67 | 1.73 |
| Acetate ($CH_3COO^-$) | 4.82 | 6.4 | 2.58 | 3.19 |

Another important observation from the simulation results is the large difference of the OAm adsorption energy for Cu- and S-rich ionic cross-sections of the same (100) facet. For the S-terminated (100) facet, the adsorption energy is nearly 2.5 to 3.9 times smaller than that for the Cu-rich atomic cross-section (options #1 and #2 compared to the option #3 from Table 1.). Hence, the passivated by OAm Cu-rich facet presents large potential barrier for the $[CuS]_x$ monomers access and hampers the growth. Obviously, an excessive amount of S-precursor in the solution in the comparison to Cu-precursor promotes the addition of monomers and further growth in kinetic mode facilitating the formation of triangles.

We assume that the most reactive sites are the vertex (tip) positions of the triangular NC, and under the sufficient supply of sulfur these positions are occupied, promoting the triangular shape of the initial NCs. Considered phase 1 occurs in a kinetic (non-equilibrium) regime, the growth is accompanied by the maximization of the surface area. After a certain period when the monomer solution is depleted, the growth changes to the thermodynamic mode. Under equilibrated monomer supply some particles grow and develop all six facets to minimize the surface energy of the system in expense of smaller triangles which slowly dissolve. Sharp tips dissolve first having the largest curvature radius and being less stable in comparison to the edge according to the Ostwald ripening process [52]. The role of sulfur anions can be compared with the role of surface passivator for Cu-rich facets, which competes with OAm promoting the observed crystal growth.

The atomic structure of triangular NC is presented in Fig. 4 C. Reactions of monomers at positions near and at the vertex (least stable, black arrows) is the vital process for retaining the triangular shape and a sufficient monomer supply rate is of great importance here. During phase 2 when the monomer supply drops substantially, the growth rate over the perimeter of the triangular NCs becomes non-homogenous. Monomers which condense at side facets obviously occupy either central positions of the edge (Fig. 4 C, orange arrows) allowing the development of other three facets from the corner (leading to truncated triangles and hexagons).

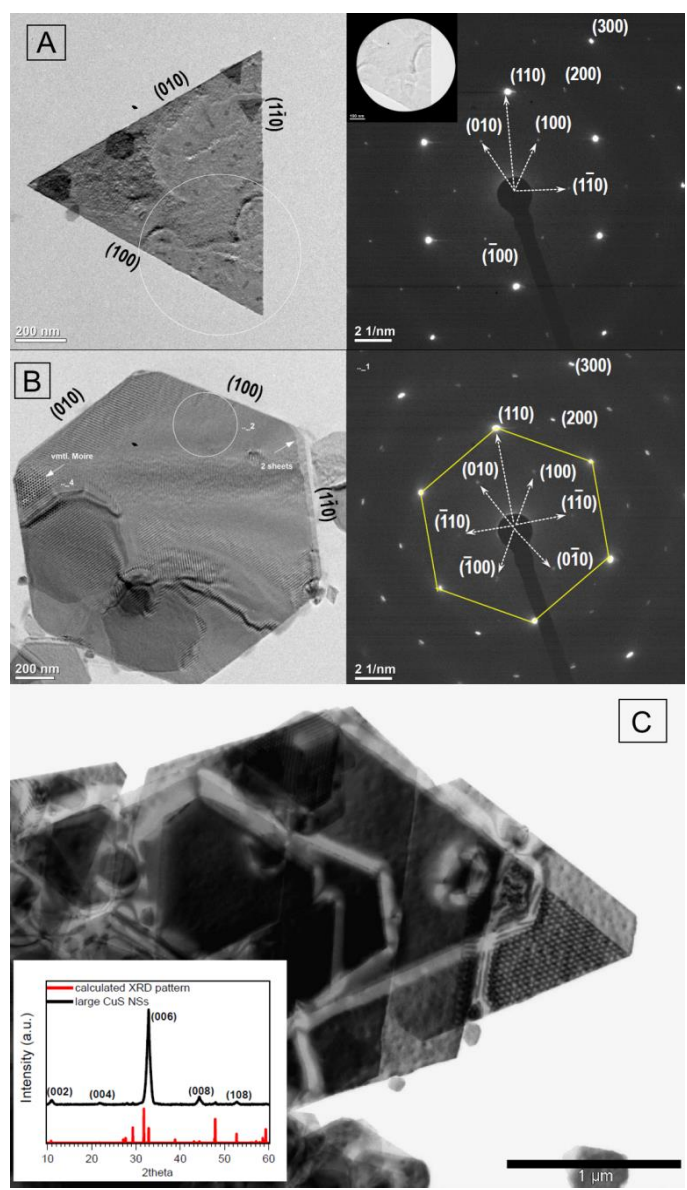

**Figure 3.** TEM image of large CuS NSs and SAED pattern with [001] zone axis: (A) Triangular NS; (B) Hexangular NS. Study of SAED patterns reveals the (100), (010), (1$\bar{1}$0) facets of CuS NSs; (C) Truncated triangular and hexangular NSs. Inset: XRD patterns of large NSs on the substrate. The texture effect influences the intensity distribution, (00$l$) reflexes become dominant, whereby the intensity of the (101), (102) and (103) reflexes drastically decreases. According to the Scherrer equation, the average thickness was found to be about 9.6 nm.

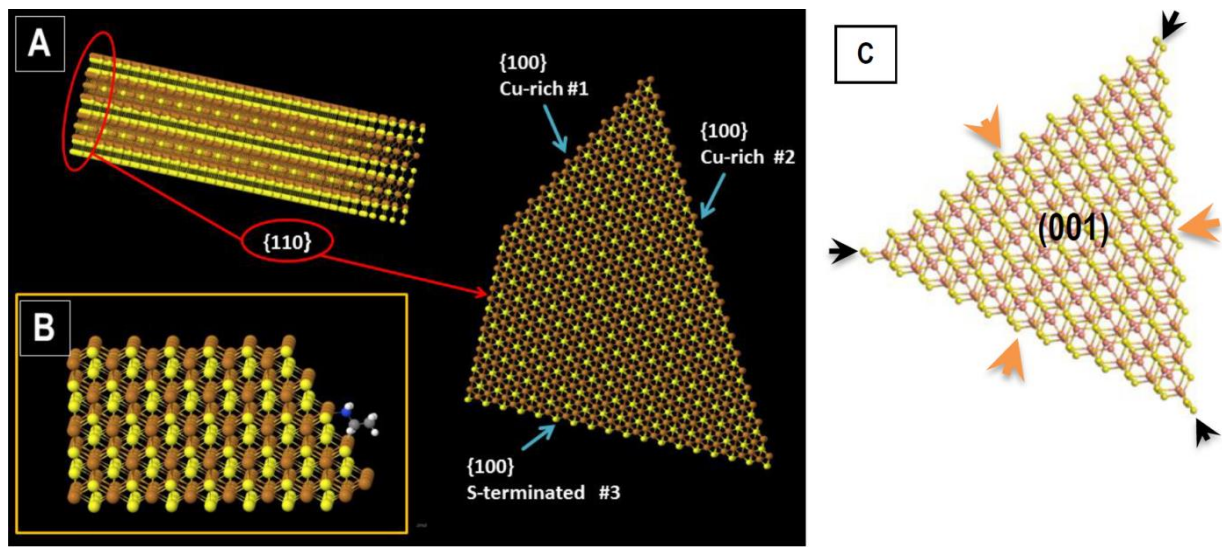

**Figure 4.** Scaled-down atomic models for CuS NCs terminated with different facets: (A) Faceting of the CuS NC according to the simulated surfaces (Table 1); (B) Exemplarily shown amine group adsorbed at {100} facet (#2) (simulation); (C) triangular CuS NC terminated at edges with sulfur atoms, vertex and central positions are denoted by arrows. The lateral crystal plane corresponds to the (001) facet.

To better understand the shape transformation of CuS nanoprisms under different conditions we investigate the halogen-free reaction, where CuI is substituted with copper acetate. For comparison, the synthesis with copper oleate is also studied, where to the reaction flask with copper acetate corresponding amount of oleic acid was added. Under vacuum (2h) and moderate heating (80°C), acetate can be transferred to oleate which has been already shown to increase the reactivity of lead cations [46]. The TEM images of two representative routes are shown in Fig. 5. Triangular and hexangular nanoprisms in this protocol adopt the same crystal orientation in respect to the shape shown by HRTEM analysis (Fig. S4).

During phase 1 in the halogen-free synthesis, the particles still tend to adopt the triangular shape (Fig. 5 A). However some irregular and hexagonal shapes are also observed as minor fraction. The short acetate group still binds strongly to the copper ions, reducing its reactivity. Transfer to the copper oleate precursor increases the reactivity of the metal. By keeping other conditions constant, we observe the adoption of both hexagonal (dominant) and triangular shapes after 1 minute after the injection of sulfur (Fig. 5 B). After 60 minutes of synthesis $Cu_xS$ nanoprisms typically transform to the polygonal particles with an average diameter of 63 nm (dispersity D=15%). The addition of precursors during phase 2 can increase this size up to 250 nm (not shown here).

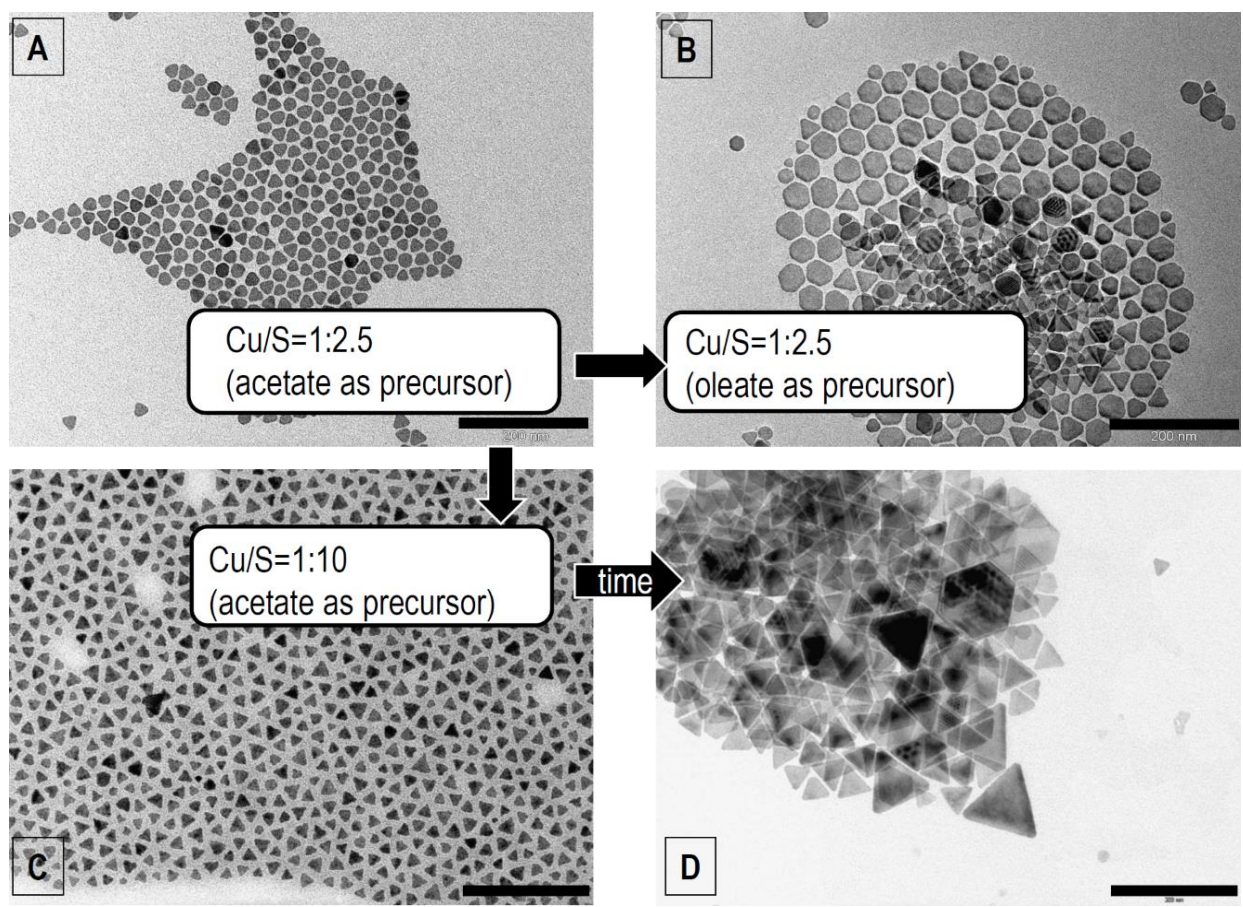

**Figure 5.** Shape changes of Cu$_x$S NCs during hot-injection synthesis in OAm. (A,B,C) TEM images of samples after 1 min synthesis; (D) after 5 minutes. (A,B) Cu:S ratio 0.4:1, (C) 0.1:1. Preparation of faceted triangles (picture C) with an average side length of 20 nm (σ=19%). The size increased with time to ~50 nm (D). Scale bars 200 nm.

If the amount of copper is reduced to a Cu:S ratio of 0.1:1 only triangles with sharp vertices are formed in the acetate-based synthesis (Fig. 5 C) with an average size of 20 nm (D=19%). During the first 5 minutes the triangles retain their shape and can grow up to a lateral size of 50 nm (D=12%), however a minor fraction of bigger triangles is present (around 100 nm). After 5 minutes, the system reaches phase 2, where size and shape transformations lead to hexagonally shaped ripened particles. If we keep the system in phase 1 (kinetic growth) by drop-wise addition of monomers with predefined Cu:S ratio, the triangular shape will be dominant even after 60 minutes of synthesis (Fig. S5).

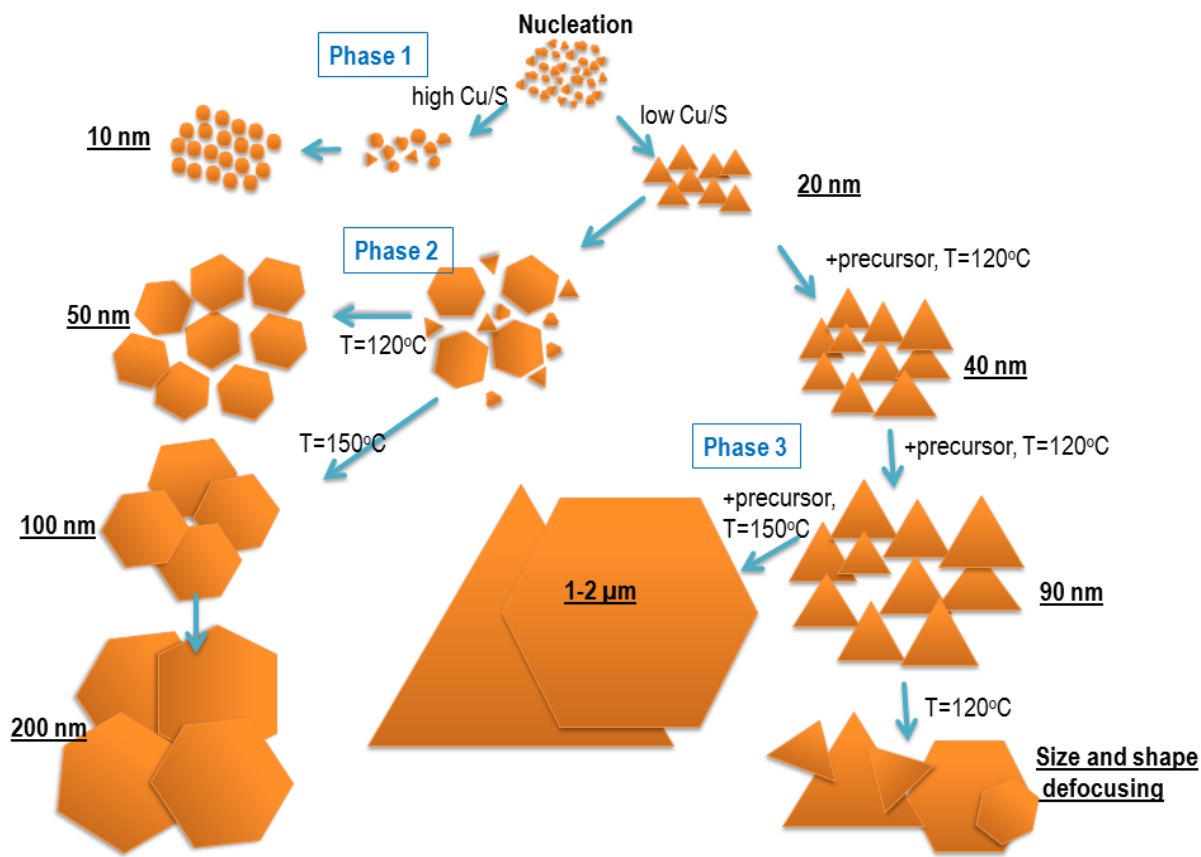

**Figure 6.** Scheme of the shape transformations during OAm-S based hot-injection Cu$_x$S synthesis depicting ways to control the shape. Blue arrows denote time flow.

The generalized mechanism for our colloidal 2D CuS growth is illustrated in the scheme in Fig. 6. It should be noted that the presence of the acetate group on the surface after the nucleation of CuS NCs might also have an impact on the shape of NCs in respect to their crystallographic orientation. However, simulations show that the adsorption energy of acetate obey a similar tendency like the amine group: the binding affinity to the Cu-rich (100) facet is at least twice larger than that for the ($\bar{1}$20) facet implying the preferential growth of edges along the {100}-like facets and not along the {$\bar{1}$20}-like facets. Due to the fact that OAm is the solvent in the system, the gradual exchange of acetate to OAm on the NC's surface is expected according to the law of mass action.

Shape control for copper sulfides was demonstrated and discussed by van der Stam *et al*. for digenite [38] and Hsu *et al*. [37] for covellite phases of copper sulfide. In the case of the rare tetragonal digenite crystal structure, the mechanism of shape transformation in fact might be based on selective passivation of specific facets by TOPO. However, in the covellite case, as shown above, this possibility is restricted due to the symmetry of the crystal, and the detailed mechanism of the halogens influence is still a challenge. However, the control over the shape for covellite can be better understood in the framework of (un)balancing the reactivities of the copper and sulfur monomers demonstrated above. Halogens are distinct nucleophiles, able to bond Cu$^+$ cations and thus, reducing the metal reactivity and shifting the balance towards sulfur. This promotes focusing of the shape to the triangular one, where halogens serve as mediators of the process.

***Optical properties of CuS nanoprisms in the NIR.*** The demonstrated synthetic tunability of the shape and size of CuS nanoprisms opens broad possibilities for the optical absorption control in the NIR. Several earlier studies reported on measurements and simulations of extinction spectra for CuS NCs [9, 19, 53]. However, the analysis for the triangular shape of CuS is still missing. In this section, we report on the optical features of 2D CuS nanoprisms with specific shape and compare them with the experimental ones. We perform our study based on the assumption that optical properties of the covellite phase in the NIR can be precisely modelled in the framework of the Drude-Sommerfeld theory. The intrinsic metallic nature of covellite shown by ab-initio simulations [31] allows one to construct the complex dielectric function according to the known relation (1) and to simulate the plasmon resonance conditions for these particles:

$$\varepsilon(\omega) = \varepsilon_\infty - \frac{\omega_p^2}{\omega(\omega+i\gamma_0)}, \qquad (1)$$

where $\varepsilon_\infty$ describes the ability of bound electrons to contribute to the polarizability, $\omega_p$ denotes the plasma frequency of CuS, and $\gamma_0$ defines the electron relaxation time (the width of the plasmon frequency in eV). The high value of the plasma frequency implies a high concentration of free carriers (holes) and defines the position of the plasmon resonance for a material. For our considerations, we use 4.0≤ $\omega_p$≤4.4 eV, $\varepsilon_\infty$=8.0 and 0.2≤ $\varepsilon_\infty$≤0.3 eV, which are in close agreement with values reported by Xie *et al.* [19] and satisfactorily reproduce our experimental spectra. Since the particles have not trivial shape we use the Discrete Dipole Approximation approach (DDA) [54], and DDSCAT 7.3 code for the simulations [55].

Figure 7A shows the spectroscopic response of the CuS NCs to the shape transformations during synthesis phases 1 and 2 (curves 1–4). In the UV-VIS region the fundamental absorption can be found below 600 nm which hardly depends on the size of particles and shows the relatively large value of the optical band gap of the synthesized material (over 2.1 eV, suggesting the covellite phase in confinement [4, 56]). The NIR region shows distinguished spectral features (LSPR resonances) essentially sensitive to shape and size changes. Initially we form triangles and allow them to grow without precursor addition in the non-kinetic mode (Fig. 6, phase 2). As discussed above, the system goes through the dissolution of smaller triangles (first of all – tips) and the formation of larger hexagonal NSs (see Fig. 1, D–G). This can be exactly tracked by absorption spectra for different temporal aliquots. The NIR spectrum for the aliquot with a major morphological yield in triangular shape (average side length of 28 nm with 4.6 nm average thickness) contains an intense broad nonsymmetrical peak centered at 1650 nm with shoulder(s) at the short-wavelength side in the 800–1200 nm range, which is typical for a triangular nanoprism localized surface plasmon resonance (LSPR) spectrum (see simulation section, Fig. 7 D). When the system goes through size and shape defocusing, tips of triangles dissolve, giving rise to the rounded shape. Therefore, the main peak shifts to lower wavelengths becoming narrower and more symmetric. This corresponds to the spectrum of thin nanodisks shown further in Fig. 7 D. At the same time, large hexagonal NSs form due to the equilibrated slow growth, giving rise to the absorption at longer wavelengths, which correspondingly can be observed in the measured spectrum (Fig. 7 A, curve 3). Eventually small particles completely dissolve and the narrow plasmon resonance peak vanishes. For larger hexagonal NSs the main (in-plane dipolar) resonance shifts further to longer wavelengths, which can be also seen in the simulations in Fig. 7 D and the spectrum in the range 1–2.5 µm contains only higher and out-of-plane modes [57] between 0.85 and 2 µm.

Applying synthetic procedures according to the scheme in Fig. 6 allows separately producing triangular and quasi-hexagonal CuS nanoprisms with controlled size in the range of about 13–100 nm (Fig. S6). Corresponding absorption spectra are displayed in Fig. 7 B, C (additionally Fig. S7) for the set of triangles and hexagon-like nanoprisms of CuS. These spectra contain similar features and demonstrate same trends in respect to shape and size compared to our simulations (Fig. 7 D, E, F). From our simulations we see that the extinction spectra of equivalent volume hexagonal, cylindrical and triangular NCs differ qualitatively and quantitatively. The peak of LSPR for spherical NCs is blue-shifted and contains one single resonance, therefore the peak is symmetric and narrow. Triangular NCs show more resonances which are successfully resolved by spectroscopy as well. In particular, the main peak is non-symmetric with shoulder(s) at shorter wavelengths, hence it is much broader. Hexangular NCs show spectral peaks between the ones of spherical and triangular NCs with a shoulder at short wavelengths (Fig. 7 E).

We note that LSPR-induced absorption of synthesized CuS NCs covers a broad range of the NIR spectral region demonstrating huge potential for applications in catalysis, NIR energy harvesting and transfer. Furthermore, the crystallization of CuS in a triangular shape theoretically gives the additional possibility to induce an substantial E-field enhancement on the tips due to the increased surface curvature [58, 59]. In Fig. 8 A we visualize the electric field intensity $|E|^2$ normalized to the incident electric field intensity (near-field mapping) for CuS nanoprisms of different shape. In an ideal case, when the tip is sharp, the triangular shape allows the field intensity to be increased by a factor up to 1400–2450 at the hot spots for the normal (in respect to the lateral surface) incidence of linearly polarized radiation. Corresponding visualization for hexagonal nanoprisms and disks with equivalent volume is shown in Fig. S8. Maximal local field intensity enhancement averaged over 75 spatial orientations of the target reaches a factor 37, 67 and 249 for disk, hexagonal and triangular prisms of equivalent volume correspondingly, making the triangular shape particularly attractive in respect to non-linear processes and energy transfer.

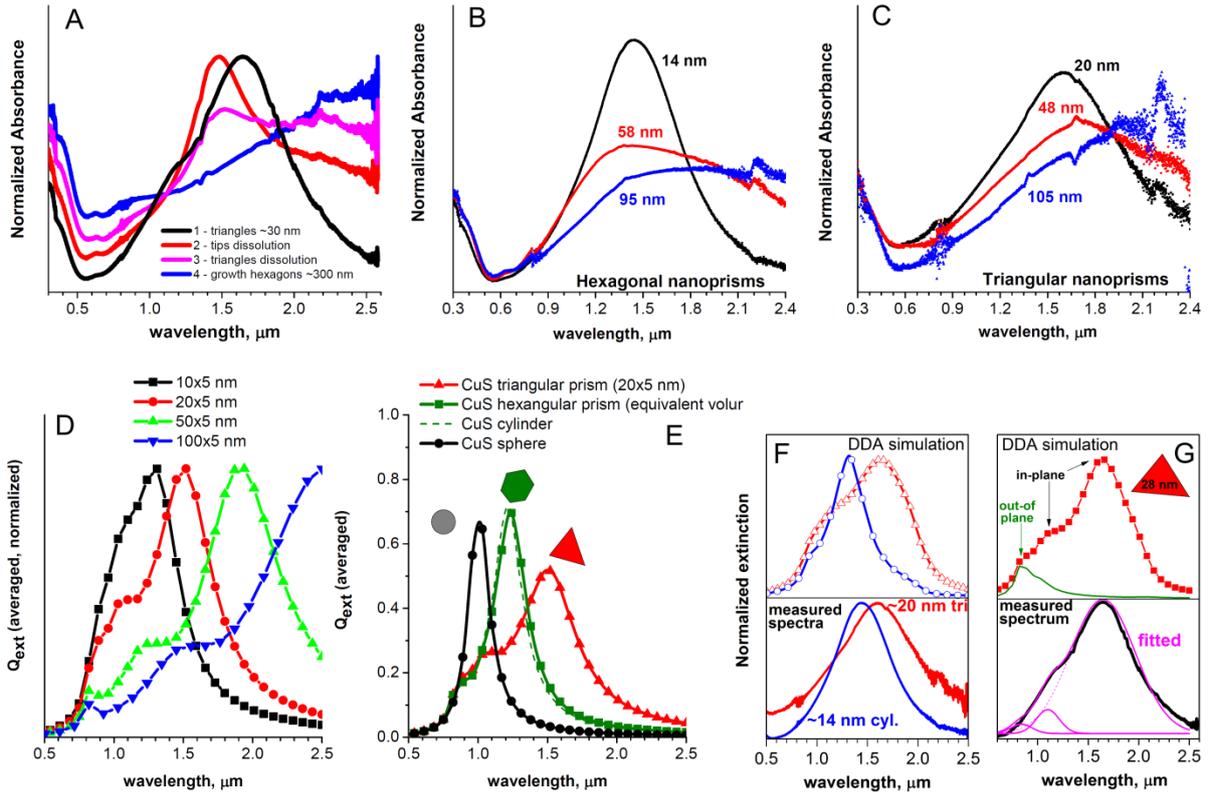

**Figure 7.** UV-VIS-NIR spectroscopic characterization of synthesized CuS NCs supported by DDA simulations: (A) Temporal evolution of the absorption spectrum of CuS NCs during the synthesis through size and shape transformations described in the main text. Primarily, small triangular nanoprisms of ~30 nm appear (curve 1), with subsequent dissolution of the tips (curve 2) and dissolution of former triangles (curve 3) during the large hexagonal NSs grow in non-kinetic mode and start to dominate; finally only the large hexagonal NSs (~300 nm) remain in the product of the reaction (curve 4). Simulated by the DDA method spectra for these particles are presented below in (D) and (E). (B) Absorption spectra of quasi-hexagonal nanoprisms (rounded side facets) prepared separately with sizes between 13 and 95 nm. (C) Absorption spectra of triangular nanoprisms (20–105 nm). (D–E) Simulated extinction spectra of CuS NCs with different shapes and sizes. Here, the results were averaged over 15 spatial orientations ($\omega_p$ =4.4 eV, $\varepsilon_\infty$=8.0, $\gamma_0$=0.2). (D) Triangular nanoprisms with the edge length of 10, 20, 50 and 100 nm and a thickness of 5 nm. (E) Comparison of the extinction spectra of triangular prism (edge 20 nm, thickness 5 nm) with hexagonal prism (thickness 5 nm), disk (thickness 5 nm) and sphere of CuS with equivalent volume (15600 dipols). (F) Comparison of the measured LSPR peak of smallest CuS NCs (disks, base diameter 14 nm, thickness 4 nm and triangles 20x4 nm) with a simulated spectrum in NIR. (G) Comparison of the measured LSPR peak (black curve) with simulated one (red) for 28x4.6 nm triangular nanoprisms with the assignment of out-of-plane resonance (green curve) and corresponding mathematical fitting of the experimental spectrum with three Gaussian curves centered at resonances positions found by the simulations (pink curves).

To assign the resonant features in the absorption spectrum of exemplarily shown spectra for triangular CuS nanoprisms in Fig. 7 G, we simulate the interaction of NCs with NIR radiation with different spatial orientations of the NCs, inducing separate out-of-plane and in-plane modes. We also map the E-field intensity at different wavelengths for qualitative assessment of the oscillating mode. The main measured LSPR resonance of 28 nm large CuS nanoprisms is situated at 1650 nm and can be attributed to the dipolar in-plane mode due to the highest intensity and sensitivity to lateral dimensions. From the simulations, it is recognizable that triangular CuS nanoprisms can produce additional resonances on the short-wavelength side of the main plasmon peak (situated around 1120 nm and 830 nm). If the polarization of the incident radiation excites only the out-of-plane resonance (incident polarization of E-field was set orthogonal to the lateral base surface), merely a broad peak at 830 nm appears (green curve in Fig. 7G) which becomes smeared out after the averaging procedure. This peak should be attributed to the out-of-plane LSPR resonance mode. We also note that this peak hardly shifts when the lateral size of particles increases drastically. Simulations with values of $\omega_p =$ 4.4, $\varepsilon_\infty$=8.0 and $\gamma_0$=0.2 show an additional weak out-of-plane peak (keep tracking green curve in Fig. 7G) at the longer wavelength side around 1025 nm. According to previous studies with silver nanoprisms and gold nanorods [60-63], the quadrupole mode appears at higher frequencies in comparison to the dipolar one, hence we can attribute the peak at 1025 nm to the dipolar out-of-plane resonance, and the peak at 830 nm to the quadrupole out-of-plane LSPR mode. We conclude that other peaks at 1120 nm and 1650 nm should be attributed to the in-plane modes. The intensity mapping of the electric field is shown in Fig. 8 for different wavelengths and polarizations of incident light on the triangular prism. The most intense peak of the spectrum of this NC (1650 nm) in the planar slice produces the field distribution which is inherent to the dipole resonant oscillations showing point-like field enhancement exactly at the tips. In contrast to this, the intensity distribution for the 1120 nm radiation appears at the sides of the NC allowing us to conclude about the quadrupole mode of the in-plane resonance. Hence, these modes should be sensitive to the lateral dimensions of the NC. In fact, we see the red-shift of the shoulder in the measured and simulated spectra. Using the values of the LSPR positions from our simulations, we satisfactorily reproduce the observed spectrum with three Gaussian curves (Fig. 7 G, pink curves) centered at 840, 1100 and 1650 nm. As nimbly noted by Xie *et al*. [19], the out-of-plane modes can be additionally influenced (dampened) by the anisotropic conductivity of covellite [23], causing less intense red-shifted resonance in the spectrum in comparison to the one predicted by the model. For smaller CuS NCs (13–20 nm, Fig. 7 F), we find that the best fit of the theoretical spectra can be performed when the plasmon frequency is reduced to 4.0 eV, with $\gamma_0$=0.25. This can be rationalized by the fact that these particles have larger surface-to-volume ratio. A larger amount of surface interaction (oxidation) and defects will reduce the effective concentration of free holes and induce the red-shift and slight broadening of the LSPR peak.

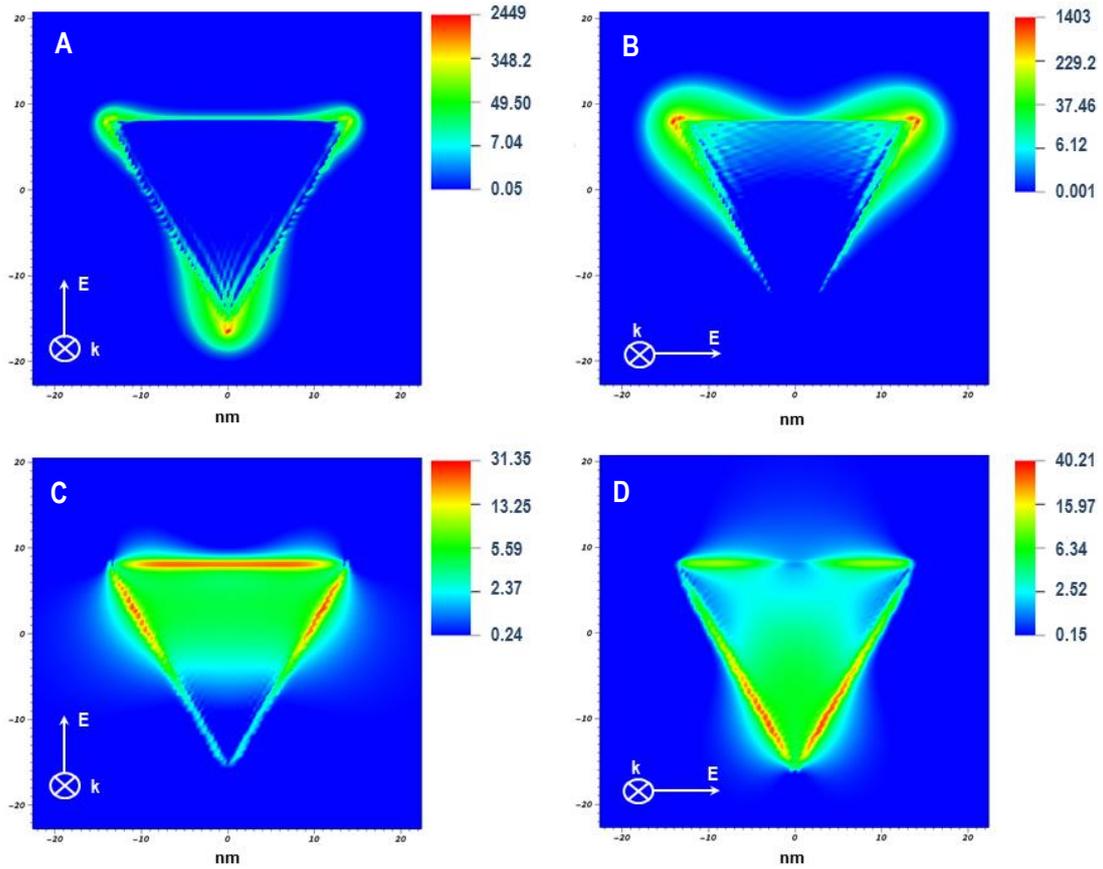

**Figure 8.** Visualization of the near-field intensity ($|E|^2/|E_0|^2$) in triangular CuS nanoprisms. The orientation of k- and E-vectors is denoted in the images. (A,B) In-plane intensity distribution at 1650 nm (dipolar LSPR mode). (C,D) In-plane intensity distribution at 1020 nm (quadrupolar LSPR mode). Because of the very high spatial field concentration at 1650 nm, a logarithmic color scaling has been applied.

It is worth to note that the position of the LSPR peak is governed by several factors, besides the geometry of the NCs. First, it is directly dependent on the plasma frequency which in turn is defined by the concentration of free carriers in the materials and the effective mass. Respectively, the surface chemistry, surface defects and minor shifts of the stoichiometry of the NC can essentially influence the effective free carrier concentration in the particle [29]. Secondly, the mutual arrangement of particles during the measurement can cause an essential red-shift of the LSPR peak in the case of side-to-side stacking or the peak broadening and damping in case of sandwich-stacking [53]. Eventually, the solvent dielectric constant plays an important role for the LSPR peak position as well. In our case the copper sulfide NCs synthesized from iodide and acetate precursors in OAm possess covellite crystal structure and in general they show a slightly shifted LSPR resonance to longer wavelegths in comparison to reported ones by e.g. Tao *et al.* or Xie *et al.* [19, 37]. However general features and NIR spectral trends of the system indicate the metallic character of the synthesized NCs (which can be true only for stoichiometric CuS) and the observed differences can be explained by the above-mentioned complex of factors influencing the LSPR.

**Conclusions**

We describe a straightforward and flexible shape control over copper sulfide nanocrystals (covellite) by balancing the precursors' reactivity and the choice of the growth regime. In our approach, we do not use additional ligands keeping the surface of the particles in its native state which comes from the synthesis conditions. We present a scheme for a one-pot synthesis which provides the possibility to produce triangle and hexagonal-like 2D nanocrystals in a controllable way in the size range between 13 and 100+ nm. We show also the possibility to grow CuS nanoprisms to micron-sized nanosheets with thicknesses of 20 nm. The main factors for the shape control and shape transformation are precursor concentrations and their ratio, temperature, time and the growth mode. By means of TEM imaging, selected area electron diffraction and DFT simulations we explain the reasons of the distinct crystallographic orientations in respect to the shape of the CuS NCs and we rationalize the importance of the sulfur-source excess during the synthesis of NCs in oleylamine for the triangular shape. Additionally, we analyze the absorption spectra in the NIR region with the help of DDA simulations and we show the influence of the shape of CuS nanoprisms on LSPR features. We confirm that the NIR optical properties of our system (intense plasmon resonance, its dependence on geometry of NC) can be well described by the Drude-Sommerfeld theory for metals. By our calculations we demonstrate huge potential of triangular CuS nanoprisms for spatial and spectral electric field concentration on the tips of the NCs in the NIR region that can serve as important material for enhanced catalytic and non-linear effects.

**Acknowledgments**

A. Kornowski, D. Weinert (HRTEM), A. Barck (XRD), L. Heymann, Dr. V. Lesnyak, Dr. C. Strelow, P. Witthöft and F. Li are acknowledged for the useful discussions. The authors gratefully acknowledge financial support of the European Research Council via the ERC Starting Grant "2D-SYNETRA" (Seventh Framework Program FP7, Project: 304980). C.K. thanks the German Research Foundation DFG for financial support in the frame of the Cluster of Excellence "Center of ultrafast imaging CUI" and the Heisenberg scholarship KL 1453/9-2.